\documentclass[preprint, prx]{revtex4-2}
\usepackage{graphicx}
\usepackage[normalem]{ulem} \usepackage{xcolor}
\usepackage{amsmath}

\begin{document}

\title{Primordial Media: the shrouded realm of composite materials}

\author{Viktor A Podolskiy}
\email{viktor\_podolskiy@uml.edu}
\affiliation{Department of Physics and Applied Physics, University of Massachusetts Lowell, Lowell, MA 01854 USA}

\author{Evgenii Narimanov}
\email{evgenii@purdue.edu} 
\affiliation{Department of ECE, Purdue University, West Lafayette, IN 47907 USA}


\date{\today}

\begin{abstract} 
Electromagnetic composites (metamaterials) recently underwent explosive growth fueled, in part, by advances in nanofabrication. It is commonly believed that as the size of the components decreases, the behavior of a composite converges to the response of a homogeneous material. Here we show that this intuitive understanding of the electromagnetic response of composite media is fundamentally flawed, even at the qualitative level. In contrast to the well-understood local effective medium response, the properties of nanostructured composites can be  dominated by electromagnetic nonlocality. We demonstrate that the interplay between the nonlocality and the structural inhomogeneity introduces two fundamentally new electromagnetic regimes, primordial metamaterials and nonlocal effective medium. We develop an analytical description of these regimes and show that the behavior of metamaterials in the limits of vanishing nonlocality and of vanishing component size do not commute. Our work opens a new dimension in the design space of nanostructured electromagnetic composites. 
\end{abstract}

\maketitle

\section{Introduction}
Composites with engineered optical response enable novel approaches to  imaging, sensing, communications, and quantum engineering\cite{quantumMM,pendrySuperlens,enghetaReview,kivsharReview,shalaevNIMreview,narimanovReview}. In order to minimize the artifacts related to light interference and scattering on an individual component and to make the composite overall resemble homogeneous media\cite{miltonBook,podolskiyNonlocalAPL}, the feature size of recent composites approach few nanometers  scale \cite{nanoHecht,nanoBaumberg,coreShellHalas,coreShellScience,wegenerReview,baumberg2019extreme,ballisticOL}.
Recent studies indicate, however, that the properties of nm-scale inclusions begin to deviate from their bulk response due to electromagnetic nonlocality\cite{stockmanImperfectLens,schatzPNASplasmonic,deAbajo,norlanderPRB}. Here we demonstrate that the novel interactions between nonlocal components of the composite yield qualitative -- not just quantitative change. Our analytical results, illustrated here on the example of planar metamaterials, demonstrate the existence of two fundamentally new electromagnetic regimes, primordial metamaterials, and nonlocal effective medium. 

\begin{figure}[b]
\centering\includegraphics[width=9cm]{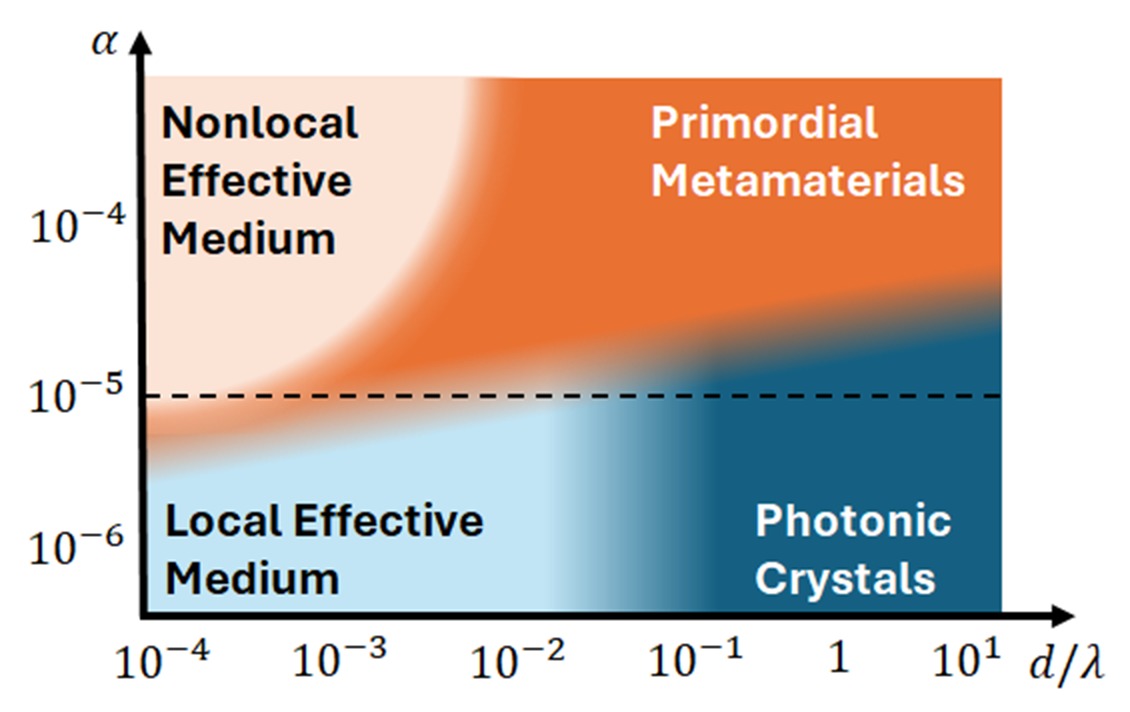}
\caption{The interplay between the space inhomogeneity (parameterized  by the composite lengthscale $d$, normalized to the vaccuum wavelength $\lambda$) and the nonlocality (parameterized by a dimensionless parameter $\alpha$ -- see the discussion following Eq.~(\ref{eq:D})) results in the complex optical response of the composites. When the nonlocality is sufficiently weak and can be ignored, the composites operate either in {\it photonic crystal} regime, dominated by interference of partially reflected light or in {\it local effective medium} regime where the system  behaves as homogeneous material. Increasing nonlocality drives the composite into {\it primordial metamaterial} regime that is dominated by the interference of additional electromagnetic waves. Very fine-scaled strongly-nonlocal composites yield {\it nonlocal effective medium} regime where composite behaves as a homogeneous -- but strongly nonlocal -- media. Dashed line illustrates the parameter range  explored in this work}
\label{figScales}
\end{figure}

The universal design space of electromagnetic composites that emerges from our work is illustrated in  Fig.\ref{figScales}. The bottom part of the figure contains the two regimes that have been extensively studied over past decades\cite{shalaevNIMreview,kivsharReview,capassoNonlocalMetasurface,yuNonlocalMetasurface,aluNonlocalMetasurface,podolskiyNonlocalPRL,narimanovReview,enghetaReview} and where nonlocality is not important,  photonic crystals, and local effective medium. The upper part of the figure contains primordial metamaterials, a regime where the composite is dominated by the interference of inherently-nonlocal additional waves and that has been recently realized in experiments\cite{primordial}. The final part of the design space is the nonlocal effective medium, that has been unexplored up until now.  The two new regimes, enabled by electromagnetic nonlocality, offer new opportunities in classical and quantum  electrodynamics.

\section{Electromagnetic response of nonlocal composites}
Electromagnetic nonlocality can be attributed to the inherent motion of charges within materials,\cite{landauECM,AgranovichBook,pekarJETP}
 necessarily imposed in any media by the fundamental quantum uncertainty principle.\cite{landauPK,landauECM} 
This phenomenon, previously analyzed in macroscopic homogeneous materials\cite{kiselevJETP73,Hopfield63} and in individual nanostructures\cite{norlanderPRB,deAbajo} can be described by introducing the dependence of the dielectric permittivity on the wavevector $\vec{k}$ in addition to its dependence on operating frequency $\omega$\cite{landauECM}. This dependence raises the degree of dispersion relations describing plane waves propagating in materials, making them at least $\propto k^4$ instead of $\propto k^2$ (see Appendix for details), and thereby introducing additional electromagnetic waves. The existence of these waves reflects the fundamental difference between the inherently nonlocal response of actual materials and their
simplified  description in terms of local (frequency-dependent) permittivity and permeability.

In contrast to previous studies that mostly focused on a single homogeneous nonlocal layer or a single nonlocal nanostructure\cite{kiselevJETP73,pekarJETP,Hopfield63,norlanderPRB}, here we consider light propagation through a nonlocal composite material with spatially-dependent nonlocality. Although the new regimes of nonlocal composites presented in this work reflect the general and universal properties of nanostructured media, to better illustrate our results, we use an example of non-magnetic ($\vec{B}=\vec{H}$) weakly-nonlocal material that is homogeneous in the $xy$ plane and is inhomogeneous and nonlocal in the $z$ direction; the formalism presented in this work can be extended to other geometries of inhomogeneity or nonlocality. 

Since describing nonlocality in the wavevector domain is known to lead to complications when materials' properties depend on position\cite{AgranovichBook,WolfABC74}, we use the representation of the nonlocal behavior in the spatial domain, where materials' response is expanded in derivatives of the electric field. For materials with inversion symmetry, the lowest-possible nonlocality appears as the second derivative of electric field\cite{AgranovichBook}.  
The operator of electric energy of the material, defined as $w_e=\frac{1}{16\pi}\int \vec{E}\cdot \vec{D} d^3r=\frac{1}{16 \pi}\int \vec{E}\cdot \hat{\epsilon} \vec{E} d^3r$, has to be a self-adjoined operator. Therefore, the real space dependence of electric displacement on the electric field takes the form: 
\begin{equation}
\vec{D}=\hat{\epsilon}\vec{E}-\frac{c^2}{\omega^2} \frac{\partial}{\partial z}\left(\alpha(z)\frac{\partial E_z}{\partial z}\right)\hat{z}. 
\label{eq:D}
\end{equation}
with $\hat{\epsilon}$ being local part of material response. For illustration purposes, we limit the discussion below to uniaxial materials with optical axis along the $\hat{z}$ direction. Therefore, tensor $\hat{\epsilon}$ is diagonal with non-zero components $\epsilon_{xx}=\epsilon_{yy}=\epsilon_\perp$, and $\epsilon_{zz}$.

The dimensionless parameter $\alpha$ in the equation above describes the nonlocality of the electromagnetic response of material. At the fundamental level this inherent nonlocality is related to the motion of the free carriers or to the fundamental qunatum mechanical position uncertainty of the localized charges. The ratio of the corresponding spatial scale $l_p$ to the free space wavelength defines $\alpha=4\pi^2 l_p^2/\lambda^2$. In dielectric media whose response is driven by bound electrons, $l_p$ is typically small, below few nanometers. In plasmonic materials $l_p\sim \lambda v_f/c$ with $v_f$ being Fermi velocity\cite{landauPK,primordial} (see below). $l_p\sim 1 \mu m$ have been reported in phononic materials\cite{kiselevJETP73}. Extended $l_p$ are also expected in excitonic materials.   

Since in our geometry the material is homogeneous and isotropic in the $xy$ plane, we  look for propagating waves that have harmonic dependence [$\vec{E},\vec{D}\ldots\propto \exp(-i\omega t+ik_x x)$], thereby fixing $xz$ plane as the propagation plane, and focusing on propagation of transverse-magnetic (TM-polarized) modes that have $\vec{B}\|\hat y$ as these are the only waves that are affected by anisotropy and nonlocality in our geometry (see Appendix). 

Starting from Maxwell equations in Cartesian coordinates: 
\begin{eqnarray}
\label{eqMaxwell}
\frac{\partial H_y}{\partial z}&=&i\frac{\omega}{c}\epsilon_\perp E_x,
\\ \nonumber
\frac{\partial E_x}{\partial z}-i k_x E_z&=&i\frac{\omega}{c}B_y,
\\ \nonumber
i k_x H_y&=&-i\frac{\omega}{c}D_z,
\end{eqnarray}
and eliminating the electric field,  we obtain the following equation governing the $B_y$ component: 
\begin{widetext}
\begin{equation}
\label{eqBy}
    \left(\epsilon_{zz}-\frac{c^2}{\omega^2}\frac{\partial}{\partial z}\alpha(z)\frac{\partial}{\partial z}\right)
    \left[\frac{\partial}{\partial z}\left\{ \frac{1}{\epsilon_\perp} \frac{\partial B_y}{\partial z}\right\} +\frac{\omega^2}{c^2}B_y\right]-k_x^2 B_y=0. 
\end{equation}
\end{widetext}
This equation represents one of the main results of the present work.  In materials with a smooth variation of the permittivity and nonlocality, Eq.(\ref{eqBy}) can be used to calculate the position-dependent distribution of the magnetic field across the composite, with the distributions of other field components given by Eq.(\ref{eqMaxwell}). In particular, in homogeneous materials, Eq.(\ref{eqBy}) accepts plane wave solution with resulting dispersion being identical to predictions of the wavenumber-dependent permittivity formalism (see Appendix).  

For materials with step-continuous distributions of the permittivity, the requirement for Eq.(\ref{eqBy}) to have well-defined solutions [the requirement to have differentiable functions within Eq.(\ref{eqBy})] solves the decades-long puzzle of additional boundary conditions (ABCs)\cite{AgranovichBook,pekarJETP,WolfABC74} that are required to calculate the amplitudes of additional waves refracting through the interface. 

The terms inside the square brackets of Eq.(\ref{eqBy}) require continuity of $B_y$, and $\frac{1}{\epsilon_\perp}\frac{\partial B_y}{\partial z}$, which -- as seen from Eq.(\ref{eqMaxwell}) {\bf --} are equivalent to continuity of $D_z$ and $E_x$, respectively. These conditions are equivalent to boundary conditions that are imposed in the conventional (local) electromagnetism\cite{landauECM}. 

It can be shown that the expression in square brackets in Eq.(\ref{eqBy}) is proportional to $E_z$. Therefore, both $E_z$ and $\alpha\frac{\partial E_z}{\partial z}$ have to be also continuous through the interface of two nonlocal media. These two conditions represent the ABCs that are derived here from purely from macroscopic electromagnetic description -- without relying on quantum-mechanical description of materials Refs.\cite{pekarJETP,kiselevJETP73,AgranovichBook}. Further analysis reveals that when $\alpha\rightarrow 0$ on one side of the interface, the former of the two ABCs becomes redundant. 

Importantly, the set of conventional and additional boundary conditions, derived in this work, ensure continuity of the normal component of Poynting flux through the interface, defined as \cite{landauECM}:
\begin{equation}
S_z=\frac{c}{8\pi}Re(E_x H_y^*)-\frac{c^2\alpha}{16\pi\omega}\left(
E_z^*\frac{\partial E_z}{\partial z}-E_z\frac{\partial E_z^*}{\partial z}\right).
\end{equation}

\section{Effective medium behavior of composites}
The Eq.(\ref{eqBy}) can be used to derive the response of composites in the effective medium limit, when the scale of variation of material 
parameters $d\rightarrow 0$ and where the (averaged over the unit cell) fields are described by 
\begin{equation}
\label{eqByEMT}
    \left(\varepsilon_{zz}-\tilde{\alpha}\frac{c^2}{\omega^2}\frac{\partial^2}{\partial z^2}\right)
    \left[\frac{1}{\varepsilon_\perp}\frac{\partial^2 B_y}{\partial z^2} +\frac{\omega^2}{c^2}B_y\right]-k_x^2 B_y=0. 
\end{equation}
with $\hat{\varepsilon}$ and $\tilde{\alpha}$ being effective medium parameters. 

Enforcing the relatively slow (as compared with $d$) variation of continuous field components, we arrive to following expressions for the components of the effective permittivity tensor $\hat{{\varepsilon}}$ of local composites:  
\begin{eqnarray}
{\varepsilon}^{l}_\perp=\langle \epsilon_\perp\rangle, 
{\varepsilon}^{l}_{zz}=\langle 1/\epsilon_{zz}\rangle^{-1},
\label{eqEMTloc}
\end{eqnarray} 
with $\langle\ldots\rangle$ representing average over the unit cell. These results are identical to effective medium parameters derived for multilayered metamaterials in previous studies\cite{podolskiyEMT,hoffman,miltonBook}. 

However, a similar procedure applied to nonlocal composites yields:
\begin{eqnarray}
 {\varepsilon}^{nl}_{\perp}=\langle \epsilon_{\perp}\rangle, {\varepsilon}^{nl}_{zz}=\langle \epsilon_{zz}\rangle,
 \tilde{\alpha}=\langle 1/\alpha\rangle^{-1}.
\label{eqEMTnl}
\end{eqnarray}

Surprisingly, the nonlocal effective medium composite {\it does not} behave as its local counterpart with a nonlocal correction. This is one of the central results of our work. 

The fact that $\hat{\varepsilon}^l\neq \hat{\varepsilon}^{nl}$ indicates that there must be a non-trivial transition region between the two effective medium regimes. Here we identify this region as primordial metamaterials regime.  

\begin{figure}[b]
\centering\includegraphics[width=13cm]{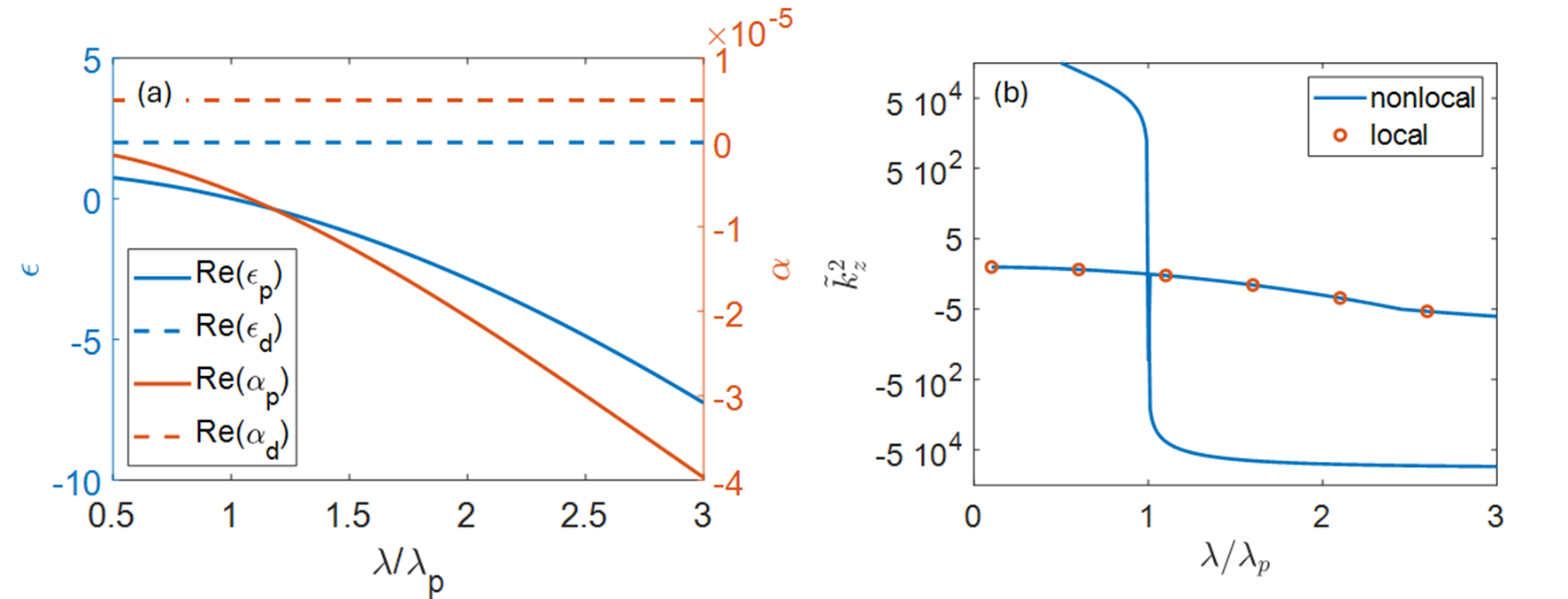}
\caption{(a) Permittivity and nonlocality of plasmonic and dielectric layers used in this work; (b)~(blue lines) Dispersion of main and additional waves in plasmonic layers; red symbols illustrate dispersion of local plasmonic materials ($\alpha=0$)}
\label{figEpsilon}
\end{figure}

\section{The breakdown of local electromagnetism in composites}
To gain an insight into the dramatic effect of weak nonlocality on optical response of composites we calculate light transmission through a stack of increasingly thinner layers. We consider a representative example of plasmonic/dielectric layers, keeping the total thickness of the stack and the concentration of plasmonic layers constant, while decreasing the size of each individual layer and increasing the total number of layers. 

We assume that the nonlocal response of plasmonic layers is well described by the permittivity of the degenerate Fermi gas\cite{landauPK}: 
\begin{eqnarray}
    \epsilon_p=1-\frac{\omega_p^2}{\omega(\omega+i\tau)}-\frac{3\omega_p^4 v_f^2 k_z^2}{5c^2\omega^2(\omega+i\tau_{nl})^2}, 
\end{eqnarray}
where $\omega_p$ is the plasma frequency, $v_f$ is Fermi velocity, and the parameters $\tau$ and $\tau_{nl}$ describe the scattering  losses in local and nonlocal regimes\cite{landauPK}, with $v_f^2/c^2\sim 10^{-5}$, $\tau=0.1\omega_p, \tau_{nl}=0.2\omega_p$. The properties of the dielectric material components are assumed to be independent of wavelength, with $\epsilon_d=2+10^{-6}(5+1i)k_z^2c^2/\omega^2$. Fig.\ref{figEpsilon}(a) illustrates permittivity of the two materials used in this study. 

In the remainder of the work we normalize all linear dimensions by the plasma wavelength $\lambda_p=2 \pi c/\omega_p$ and describe the wavenumbers using the dimensionless wavevector $\tilde{k}_\beta={k_\beta} c/\omega$ with $\beta$ representing Cartesian components. 

\begin{figure}[b]
\centering\includegraphics[width=13cm]{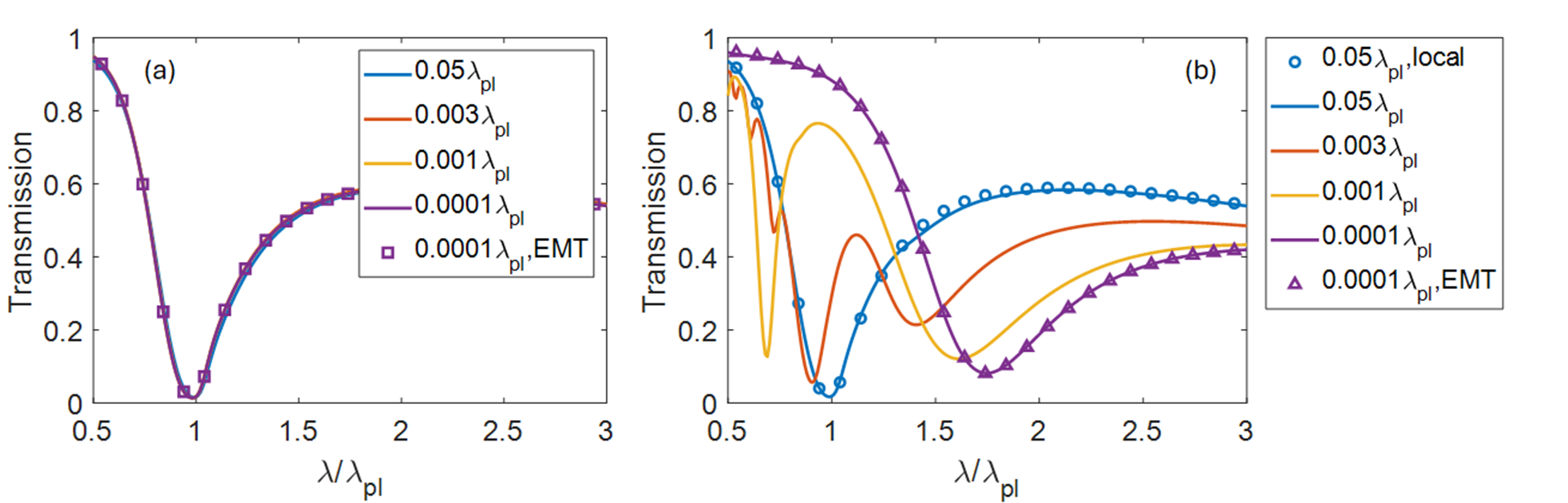}
\caption{Transmission of TM-polarized light, incident at $60^o$, through $\lambda_p/5$-thick stacks of plasmonic/dielectric multilayer composites, with different layer thicknesses $d$, calculated using (a) local and (b) nonlocal transfer matrix method, TMM [see Appendix]; squares and triangles represent results of local [Eq.(\ref{eqEMTloc})] and nonlocal [Eq.(\ref{eqEMTnl})] effective medium description, respectively; circles in (b) correspond to the calculation using local TMM.}
\label{figTransmission}
\end{figure}

The dispersion of the main and additional waves in the plasmonic component of our composites is shown in Fig.\ref{figEpsilon}(b). Note that both main and additional waves propagate for shorter wavelengths ($\tilde{k}_z^2>0$ for $\lambda<\lambda_p$) while they exponentially decay for $\lambda>\lambda_p$ ($\tilde{k}_z^2<0$). It is also seen that the dispersion of the main wave is well-described by local permittivity model ($\alpha=0$) and that outside epsilon-near-zero region  the additional wave has extremely large propagation constant and, as a result, it is not expected to couple to conventional diffraction-limited light. However, the wavenumber of the additional wave defines a new, primordial, length scale $l_p\sim \lambda| \tilde{k}_z|$ that, as we will show here, is crucial to understanding the response of nonlocal composites.  

\begin{figure*}[bt]
\centering\includegraphics[width=13cm]{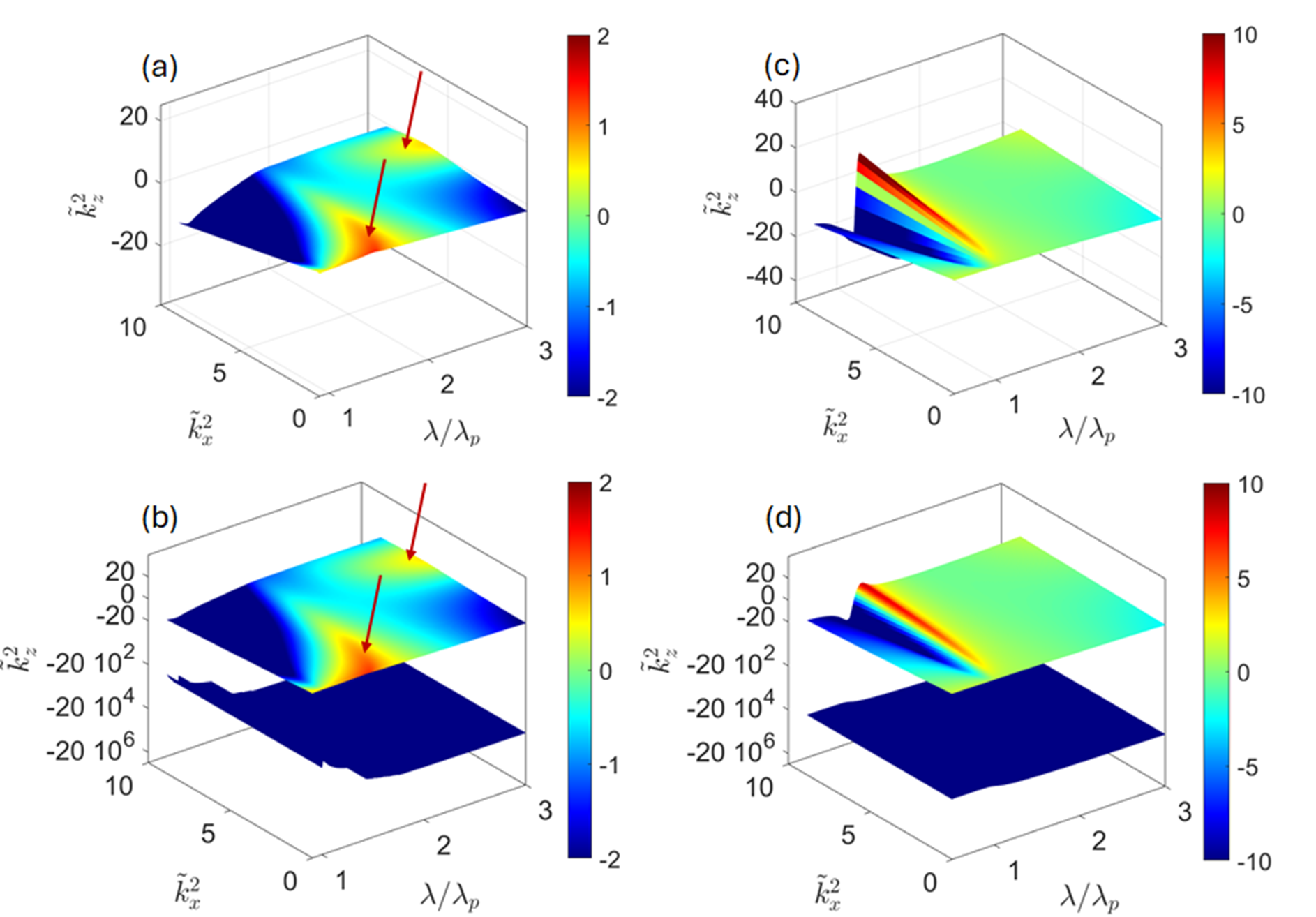}
\caption{Dispersion of the modes in periodic plasmonic/dielectric composites with layer thickness $d=0.3\lambda_p$ (a,b) and $d=0.05\lambda_p$ (c,d); $\lambda/\lambda_p,\tilde{k}_x$, and $\tilde{k}_z$ represent dimensionless operating wavelength, in-plane and out-of-plane components of dimesionless wavenumber, respectively; $\tilde{k}_z^2>0$ corresponds to the   propagating modes; $\tilde{k}_z^2<0$ represent exponentially decaying waves; panels (a,c) represent the results of local transfer matrix method and local effective medium theory, respectively; panels (b,d) represent the results of the nonlocal transfer matrix method. The two propagating modes seen in (a,b) (arrows) originate from the coupled surface plasmon polaritons\cite{Avrutsky}; the sharp resonance in (c,d) corresponds to the transition between elliptic ($\lambda<\lambda_p$) and hyperbolic ($\lambda>\lambda_p$) regimes}
\label{figThick}
\end{figure*}

Fig.\ref{figTransmission}(a) illustrates the evolution of the  transmission of TM-polarized light incident at $60^o$ through a $\lambda_p/5$-thick multilayer stack when nonlocality is ignored, and when the thickness of individual layer{\bf s} $d$ changes from $\lambda_p/20$ to $\lambda_p/10^4$. As expected, overall transmission through the composite does not depend on the (deeply subwavelength) thickness of its component{\bf s}, reflecting effective-medium-regime behavior; the dip at $\lambda\simeq \lambda_p$ indicates the transition of the effective medium response from the elliptic $\varepsilon^l_\perp,\varepsilon^l_{zz}>0$ to the hyperbolic $\varepsilon^l_\perp>0,\varepsilon^l_{zz}<0$ regime\cite{hoffman}. 

However, as seen in Fig.\ref{figTransmission}(b), this simple dynamics fundamentally changes when  the nonlocality is taken into account. When the layer thickness is sufficiently large (here, $d  \gtrsim   0.05\lambda_p$), transmission through nonlocal composite is virtually identical to transmission through its local counterpart. As the layers become thinner, the single transmission dip splits into multiple minima, one of which moves towards the longer wavelengths,  while the other shifts in the opposite direction. This is a direct manifestation of the primordial metamaterials introduced in the present work. 

Finally, when the layers reach $d\sim 10^{-4}\lambda_p$, the transmission spectrum, once again, converges to a spectrum featuring a single minimum. In this regime, the composite once again behaves as an effective medium.

\section{Evolution of modes in nonlocal composites}

To further understand the different regimes of the composites' response, we analyze the modes propagating in  periodically stratified bi-layer composites. Dispersion of these modes can be analyzed by enforcing the Bloch periodicity relationship: $\vec{E}(z+\Delta)=e^{i\tilde{k}_z\Delta\omega/c} \vec{E}(z)$, $\vec{B}(z+\Delta)=e^{i\tilde{k}_z\Delta\omega/c} \vec{B(z)}$ with  $\Delta$ being the period of the composite, resulting in the eigenvalue problem:  
\begin{equation}
    {\rm det}\left|\hat{T}_\Delta -e^{i\tilde{k}_z \Delta\omega/c}\hat{I}\right|=0, 
    \label{eqPC}
\end{equation}
with $\hat{T}_\Delta$ being the transfer matrix of one period of the composite (see Appendix).

\begin{figure*}[ht!]
\centering\includegraphics[width=13cm]{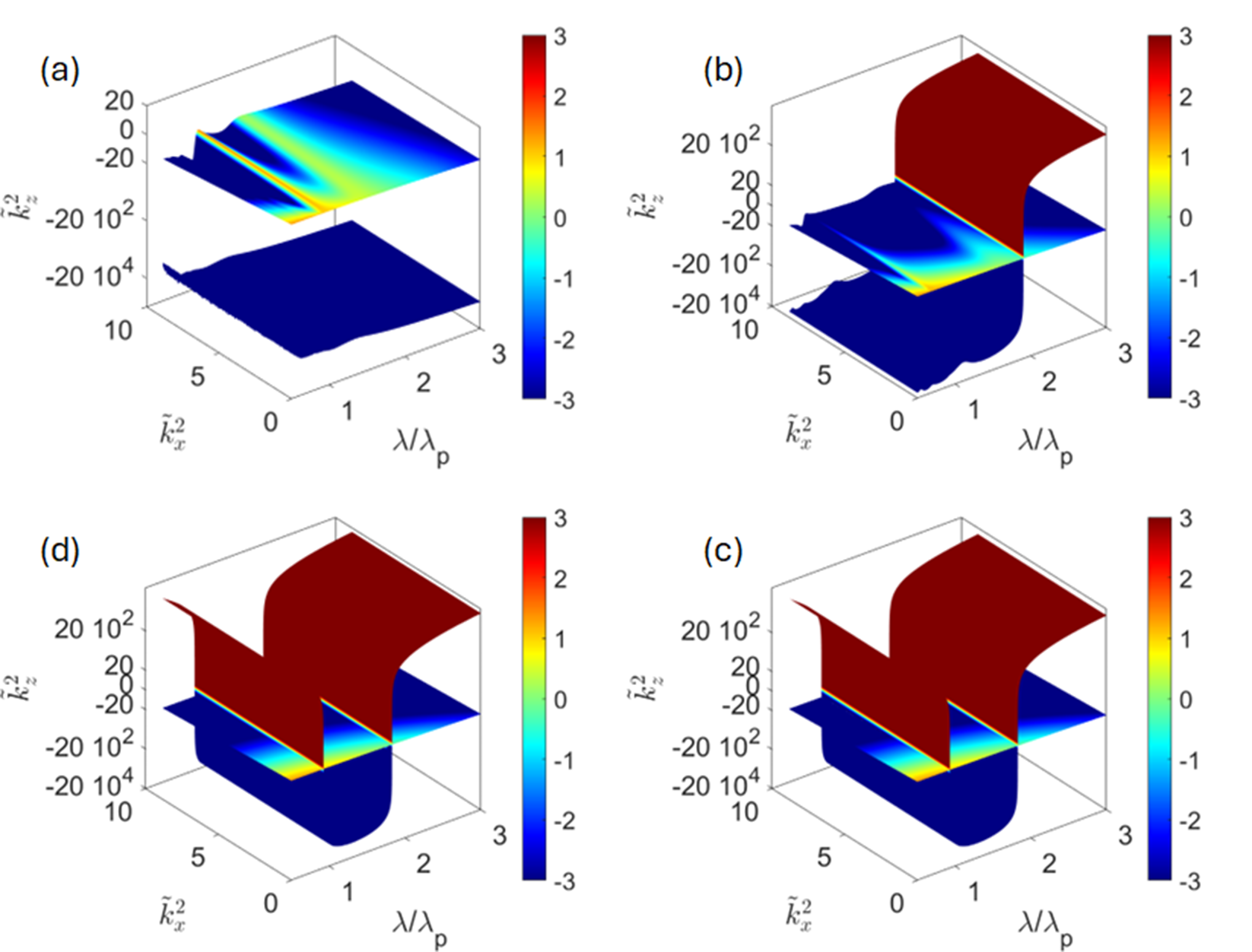}
\caption{Dispersion of the modes in nonlocal plasmonic/dielectric composites with layer  thickness $d=0.03\lambda_p$ (a), $d=10^{-3} \lambda_p$ (b), and $d=10^{-4}\lambda_p$ (c,d), calculated using nonlocal transfer matrix method (a-c) and nonlocal effective medium theory (d); notice the difference from predictions of local effective medium theory (Fig.\ref{figThick}c)}
\label{figThin}
\end{figure*}

Fig.\ref{figThick} illustrates {the} dispersion of the modes in relatively thick ($d\gg l_p$) composites, calculated using local and nonlocal descriptions. It is clearly seen that local electromagnetism adequately describes propagation of the (main) mode, with the nonlocality 
only resulting in an additional wave that exponentially decays into the composite at a very small spatial scale. 

The evolution of the material properties between $d=0.3\lambda_p$ and $d=0.05\lambda_p$ reflects the transition of the composite from photonic crystal to local effective medium regimes. In the former regime, the behavior of composite is dominated by the interference of (partially) reflected light on the scale of the unit cell, resulting in a set of transparent ($k_z^2>0$) and ``forbidden'' ($k_z^2<0$) bands; the two propagating modes shown in Fig.\ref{figThick}(a,b) corresponding to coupled surface plasmon polaritons propagating in metal-dielectric stacks\cite{podolskiyNonlocalAPL,Avrutsky}

As the layer size is reduced below the (internal) wavelength, the response of the local composite converges to the predictions of the effective medium theory, Eq.(\ref{eqEMTloc}), shown in Fig.\ref{figThick}(c,d). As mentioned above, and as described in Ref.\cite{hoffman,podolskiyPRB,podolskiyEMT,kivsharReview}, here the multilayer composite behaves as a uniaxial material with elliptic ($\lambda<\lambda_p$) or hyperbolic ($\lambda>\lambda_p$) response. From this point on, further reduction of the layer thickness does not change the predictions of the local calculations. 

However, when the layer thickness is decreased further and becomes comparable to the intrinsic component nonlocality scale $l_p$, nonlocal calculations indicate fundamental changes in composite behavior. This dynamics is illustrated in Fig.\ref{figThin}. As the layer thickness is decreased, the additional wave initially moves closer to the main mode  and begins to modulate its dispersion (Fig.\ref{figThin}a). In thinner composites, the additional wave eventually changes its behavior from evanescent to propagating (Fig.\ref{figThin}b). The behavior of the composite across this regime, {\it primordial metamaterial}, originates from the interference of additional waves in metamaterial components, with the scale of the component being comparable to the wavelength of these additional waves. As a result, the optical response of primordial composites features highly oscillatory fields\cite{primordial} and is not described by effective medium theories. In a sense, the primordial metamaterial is a nonlocal analog of the photonic crystal.

As the layer thickness is reduced further and  becomes smaller than the nonlocality scale,  the electromagnetic fields become smooth within the  unit cell. In this regime (Fig.\ref{figThin}c,d) the optical response of the composite is once again described by -- now nonlocal -- effective medium theory. The nonlocal effective medium composite supports propagation of two plane-wave-like modes, whose dispersion is given by Eqs.(\ref{eqByEMT},\ref{eqEMTnl}). As seen from Fig.\ref{figThin}d, nonlocal effective medium perfectly describes propagation of waves in very-finely-structured nonlocal composites. By comparing Fig.\ref{figThin}b to Fig.\ref{figThin}(c,d) it can be seen that the response of the composite in long-wavelegth limit homogenizes before its response in shorter wavelengths.

\begin{table*}[t]

\centering\includegraphics[width=13cm]{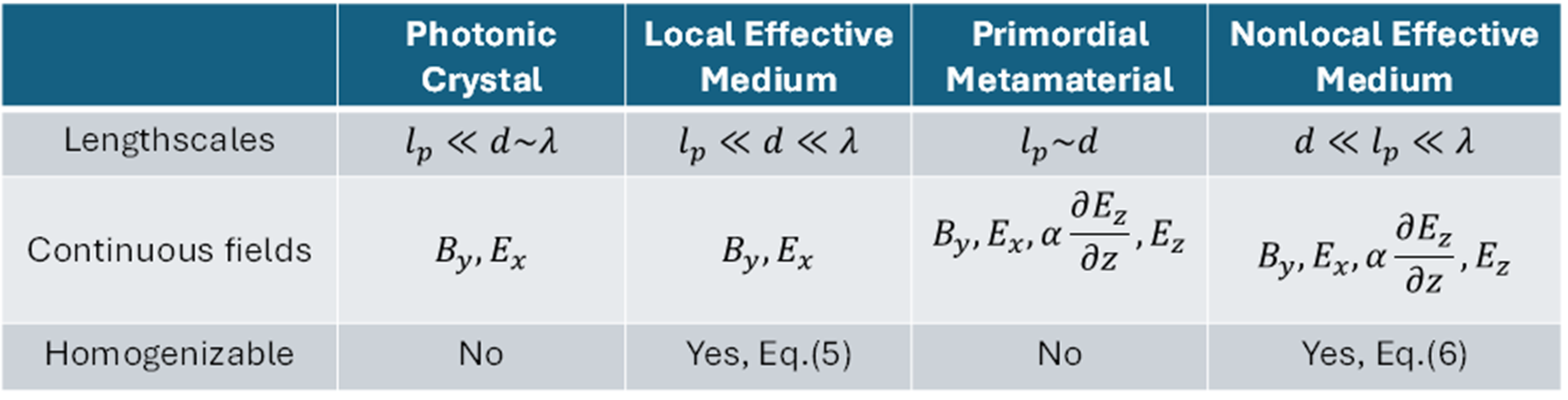}

    \caption{Different regimes of electromagnetic response of composites}
    \label{figTable}
\end{table*}

\section{Discussion}
One of the important points of this work lies in the fact that the optical properties of composites are determined not simply by the relationship between the operating wavelength $\lambda$ and the inhomogeneity scale $d$ but by complex interplay between $\lambda,d$,  and the primordial nonlocality scale ${l_p}$. The existence of $l_p$ necessitates the primordial regime and explains the fact that the limits of material response when $\alpha\rightarrow 0$ and $d\rightarrow 0$ do not commute. 

The composite operating in regime $l_p\ll d\ll\lambda$ behaves according to the local effective medium theory; the composite operating in the regime $d\ll l_p\ll\lambda$ behaves according to the rules of nonlocal effective medium theory; the regime $d\sim l_p$ represent primordial metamaterial, dominated by the interference of additional waves and whose properties, as a result, are not described by an effective medium theory -- even when the layers are very thin (see Table \ref{figTable}).

The existence of the primordial metamaterial regime reconciles the different effective medium parameters, predicted by Eq.(\ref{eqEMTloc}) and Eq.(\ref{eqEMTnl}). Indeed, when $l_p\ll d$, additional waves do not efficiently couple to the main modes. The local part of $D_z$, $\epsilon_{zz} E_z$ slowly varies across the composite, resulting in Eq.(\ref{eqEMTloc}).

In contrast, when $l_p\gtrsim d$, the displacement field is heavily influenced by nonlocal contributions, and coupling between main and additional modes cannot be ignored. When $d\ll l_p\ll\lambda$, it is $E_z$ continuity that dominates the effective medium response, resulting in Eq.(\ref{eqEMTnl}). 


As seen from Eq.(\ref{eqEMTnl}), the material with weakest nonlocality defines the overall nonlocal scale of the composite. Therefore both layers must be nonlocal to achieve primordial (or nonlocal effective medium) responses in bilayer metamaterials considered here. In the more complicated, multi-component structures, the primordial response may manifest itself through cavity resonances of additional waves, surface analogs of additional waves, etc.; these phenomena may be realized without continuous nonlocal response throughout the composite. 

Lastly, it is important to note that since nonlocality in majority of dielectric materials is relatively weak ($l_p$ is often of the order of nm),  nonlocal effective medium regime may be difficult to achieve in experiments. In contrast, primordial response should be widely available, especially in plasmonic, phononic, and excitonic materials that have $l_p\gtrsim 10 nm$. In composites based on highly nonlocal materials (where $l_p\sim\lambda$\cite{kiselevJETP73}), local effective medium regime may be not achievable with primordial response directly following local photonic crystal regime.

\section{Conclusions}
To conclude, we have developed a theoretical formalism describing electromagnetic response of composites having nonlocal components. We demonstrated that the design space for electromagnetic response of such materials contains four fundamentally different regimes of optical behavior: local photonic crystals, local effective medium, primordial metamaterial, and nonlocal effective medium. Finally, we explained the relationship between these regimes and the three lengthscales involved in the problem, layer size $d$, nonlocality scale $l_p$, and wavelength scale $\lambda$. 

Since nonlocality is an inherent property of any material, the two new regimes identified in this work, primordial metamaterial and nonlocal effective medium theory, need to be considered in understanding and engineering of optical behavior of nanostructured composites. 

Our results, illustrated here on (bi-component) multilayered metamateirals with quadratic nonlocality can be directly applied to more complicated layered media, and can be generalized to different geometries (spherical, cylindrical, etc.) and to different nonlocality responses.

This research has been sponsored by the National Science Foundation DMREF program (award \#2118787 (VP), award \#2119157 (EN))

\section*{Appendix}
\subsection{Plane waves in nonlocal uniaxial materials}
The dispersion of plane waves propagating through a homogeneous non-magnetic media can be found by substituting the plane-wave solution [$\vec{E}(\vec{r},t)=\vec{E}_0 \exp(-i\omega t+i\vec{k}\cdot\vec{r})$] into Maxwell equations and reducing these equations to the eigen value problem\cite{landauECM}. When nonlocality is described in the wavenumber domain [$\hat{\epsilon}(\omega,\vec{k})$], the above procedure yields: 
\begin{equation}
    \vec{k}(\vec{k}\cdot\vec{E}_0)-\vec{E}_0 k^2+\hat{\epsilon}\frac{\omega^2}{c^2}\vec{E}_0=0.
    \label{eqEigen}
\end{equation}
Here, the wavevector $\vec{k}$ and electric field magnitude $\vec{E}_0$ play the role of the eigenvalue and eigenvector, respectively. 

For the case of uniaxial anisotropic media with weak spatial dispersion along its optical ($\hat{z}$) axis considered in this work, permittivity is described by a diagonal tensor $\hat{\epsilon}$ with non-vanishing components $\{\epsilon_\perp,\epsilon_\perp, \epsilon_{zz}+\alpha \tilde{k}_z^2\}$. As can be explicitly verified, Eq.(\ref{eqEigen}) yields two classes of solutions. The solutions of the first class have $\vec{E}_0\perp \hat{z}$. These solutions, also known as ordinary or transverse-electric (TE)-polarized waves are unaffected by anisotropy or nonlocality and have dispersion $\tilde{k}_x^2+\tilde{k}_y^2+\tilde{k}_z^2=\epsilon_\perp$. 

The second class of solutions has $\vec{B}\perp \hat{z}$, and is therefore known as transverse-magnetic (TM)-polarized or extraordinary waves. The dispersion of these modes is given by:  
\begin{equation}
\label{eqNonlocalHom}
    \alpha \tilde{k}_z^4+(\epsilon_{zz}-\alpha\epsilon_\perp) \tilde{k}_z^2
    +\epsilon_\perp\left(\tilde{k}_x^2+\tilde{k}_y^2-\epsilon_{zz}\right)=0. 
\end{equation}
Note that $\alpha\neq 0$ yields two TM-polarized modes with different dispersion (each of these modes can propagate in either $+\hat{z}$ or $-\hat{z}$ directions). In the limit of vanishingly small nonlocality $\alpha\rightarrow 0$, the dispersion of one of these modes (main wave) approaches the well-known dispersion of the extraordinary waves in homogeneous local uniaxial media $(\tilde{k}_x^2+\tilde{k}_y^2)/\epsilon_{zz}+\tilde{k}_z^2/\epsilon_\perp=1$, while the dispersion of the second (additional) wave diverges $\tilde{k}_z\rightarrow\infty$. 

\subsection{Transfer matrix method for nonlocal multilayered media}

In optics, transfer matrix method (TMM)\cite{YehTMM} is often used to solve for light propagation in multilayered composites. Within this formalism (that can be extended to cylindrical, and other geometries), the solutions to Maxwell equations within each homogeneous layer are represented as a linear combination of (plane) waves, and boundary conditions are used to relate the amplitudes of these waves in neighboring layers via transfer matrices. 

Over the years, multiple realizations of TMM\cite{tmmReview} have been developed. Here we utilize the TMM framework to analyze light propagation through composites that include nonlocal components.  
We follow the recipe of Ref.\cite{YehTMM}, and represent the fields in each layer as a linear combination of the [TM-polarized] waves propagating in $+\hat{z}$ and in $-\hat{z}$ directions  with amplitudes $c_{l;1,2}^\pm$, with superscript defining the direction of the wave and subscripts defining the layer and the mode index ($1$=main wave, $2$=additional wave), respectively. We assume that the permittivity of the layer $l$ is given by the tensor with diagonal components by $\{\epsilon_{l,\perp},\epsilon_{l,\perp},\epsilon_{l,zz}+\alpha_{l} k_z^2 c^2/\omega^2\}$ and that the interface between layers $l$ and $l+1$ is located at $z_l$.    
We then introduce two layer-specific matrices $\hat{N}_l$ and $\hat{F}_l(z)$: 
\begin{widetext}
    
\begin{eqnarray}
\hat{N}_l=
\begin{bmatrix}
    1 & 1 & 1 & 1 \\
    \frac{\epsilon_{l,\perp}}{\tilde{k}_{z_{l,1}}} & \frac{\epsilon_{l,\perp}}{\tilde{k}_{z_{l,2}}} &
    -\frac{\epsilon_{l,\perp}}{\tilde{k}_{z_{l,1}}} & -\frac{\epsilon_{l,\perp}}{\tilde{k}_{z_{l,2}}} \\
    \alpha_l({\tilde{k}^2_{z_{l,1}}-\epsilon_{l,\perp}}) &
    \alpha_l(\tilde{k}^2_{z_{l,2}}-\epsilon_{l,\perp}) &
    \alpha_l(\tilde{k}^2_{z_{l,1}}-\epsilon_{l,\perp}) &
    \alpha_l(\tilde{k}^2_{z_{l,2}}-\epsilon_{l,\perp}) \\

    \frac{\tilde{k}^2_{z_{l,1}}-\epsilon_{l,\perp}}{\tilde{k}_{z_{l,1}}} &
    \frac{\tilde{k}^2_{z_{l,2}}-\epsilon_{l,\perp}}{\tilde{k}_{z_{l,2}}} &
    -\frac{\tilde{k}^2_{z_{l,1}}-\epsilon_{l,\perp}}{\tilde{k}_{z_{l,1}}} &
    -\frac{\tilde{k}^2_{z_{l,2}}-\epsilon_{l,\perp}}{\tilde{k}_{z_{l,2}}} 
\end{bmatrix}, \nonumber
\\
\hat{F_l}(z)=
\begin{bmatrix}
    \exp(i k_{z_{l,1}}z) &0 &0 &0 \\
    0& \exp(i k_{z_{l,2}}z) &0 &0 \\
    0& 0& \exp(-i k_{z_{l,1}}z) &0 \\
    0& 0& 0& \exp(-i k_{z_{l,2}}z)    
\end{bmatrix},
    \label{eqNnl}
\end{eqnarray}
\end{widetext}
with the columns of the first matrix representing the field distributions of individual modes in the layer, the rows representing the field components of these modes ($E_x$, $B_y$, $\alpha_l \partial E_z/\partial z$, and $E_z$, respectively), and  the diagonal elements of the second matrix representing the phase factors of the modes. 

With the notations above, the boundary conditions between nonlocal layers $l$ and $l+1$ reduce to:
\begin{equation}
    \hat{N}_l\hat{F}_l(z_l)\vec{c}_l=\hat{N}_{l+1}\hat{F}_{l+1}(z_l)
    \vec{c}_{l+1}
    \label{eqBC}
\end{equation}
where we combined the amplitudes of the (four) waves propagating in the layer $l$ into a column vector $\vec{c}_l=\{c_{l;1}^+,c_{l;2}^+,c_{l;1}^-,c_{l;2}^-\}$. 

When the layer stack contains only nonlocal components, Eq.(\ref{eqBC}) can be used to introduce the interface transfer matrix, $\hat{T}_l=\hat{F}_{l+1}(z_l)^{-1}\hat{N}_{l+1}^{-1}\hat{N}_l\hat{F}_{l}(z_l)$ that relates the amplitudes of the plane waves in the neighboring layers: $\vec{c}_{l+1}=\hat{T}_l\vec{c}_l$. 

This relationship can be used to derive the dispersion of the modes in periodically stratified materials. For the bi-layer periodic layered composite with layer thickness $d$, the overall transfer matrix of one period becomes: 
\begin{equation}
    \hat{T}_\Delta=\hat{F}_1(d)\hat{N}_1^{-1}\hat{N}_2 \hat{F}_2(d)\hat{N}_2^{-1}\hat{N}_1,
    \label{eqTDelta}
\end{equation}
with dispersion of the Bloch modes given by Eq.(\ref{eqPC}) of the main manuscript. 

For nonlocal composites with a finite number of layers, transfer matrices can be used to relate the amplitudes of waves in the first layer to the amplitudes in the last layer. The resulting relationship can then be used to solve for the amplitudes of the overall reflected and transmitted waves in therms of the amplitudes of incident waves, followed by calculation of the amplitudes of the waves throughout the layered composite.  

The TMM describing the properties of the layered stacks comprising exclusively local components\cite{YehTMM,tmmReview} can be recovered by using $2\times 2$ material matrices, 
\begin{eqnarray}
\hat{N}_l=
\begin{bmatrix}
    1 & 1  \\
    \frac{\epsilon_{l,\perp}}{\tilde{k}_{z_{l}}}  &
    -\frac{\epsilon_{l,\perp}}{\tilde{k}_{z_{l}}}  
\end{bmatrix}, 
\hat{F_l}(z)=
\begin{bmatrix}
    e^{i k_{z_{l}}z} &0 \\
    0& e^{-i k_{z_{l}}z} \\
\end{bmatrix},
    \label{eqNloc}
\end{eqnarray}
and using two-component column vector of mode amplitudes $\vec{c}_l=\{c_l^+,c_l^-\}$. 

The case of the multi-layer stacks containing both local and nonlocal components where the number of modes changes across the composite is most conveniently described using scattering- (as opposed to transfer-) matrix formalism\cite{tmmReview}. 

Within the scattering-matrix framework, the boundary conditions are written in the form given by Eq.(\ref{eqBC}), with the $\hat{F}_l(z) \vec{c}_l$ being either two- or four-component vectors (for local and non-local layers respectively), while the material matrices $\hat{N}_l$ are selected according to the following rules that reflect the different number of ABCs required at different interfaces:
\begin{widetext}
\begin{itemize}
    \item for nonlocal-nonlocal material interfaces $\hat{N}_l$ is given by Eq.(\ref{eqNnl}),
    \item for local-local layer interfaces $\hat{N}_l$ is given by Eq.(\ref{eqNloc}),
    \item for local component of the local-nonlocal interface, $\hat{N}_l=
\begin{bmatrix}
    1 & 1  \\
    \frac{\epsilon_{l,\perp}}{\tilde{k}_{z_{l}}}  &
    -\frac{\epsilon_{l,\perp}}{\tilde{k}_{z_{l}}}  \\ 0& 0
\end{bmatrix},$
    \item lastly, for nonlocal component of the local-nonlocal interface, \\$
    \hat{N}_l=
\begin{bmatrix}
    1 & 1 & 1 & 1 \\
    \frac{\epsilon_{l,\perp}}{\tilde{k}_{z_{l,1}}} & \frac{\epsilon_{l,\perp}}{\tilde{k}_{z_{l,2}}} &
    -\frac{\epsilon_{l,\perp}}{\tilde{k}_{z_{l,1}}} & -\frac{\epsilon_{l,\perp}}{\tilde{k}_{z_{l,2}}} \\
    \alpha_l\left({\tilde{k}^2_{z_{l,1}}-\epsilon_{l,\perp}}\right) &
    \alpha_l\left((\tilde{k}^2_{z_{l,2}}-\epsilon_{l,\perp}\right) &
    \alpha_l\left((\tilde{k}^2_{z_{l,1}}-\epsilon_{l,\perp}\right) &
    \alpha_l\left((\tilde{k}^2_{z_{l,2}}-\epsilon_{l,\perp}\right)  
\end{bmatrix}. 
$
\end{itemize}
    
\end{widetext}
The resulting boundary conditions are then used to solve for the amplitudes of the outgoing modes ($c_l^-, c_{l+1}^+$) in terms of the amplitudes of the incoming modes ($c_l^+,c_{l+1}^-$). On the implementation level, it is convenient to separate the resulting relationship into ``transmission'' and ``reflection'' matrices that relate $c_{l+1}^+$ and $c_{l}^-$ amplitudes to the amplitude of the incident fields, respectively, and to iteratively calculate the amplitudes of the fields, starting at the last interface of the stack (where $c_{l+1}^-=0$), and working towards the first interface of the stack. A sample code providing such an implementation can be found in \cite{githubPrimordial}.

\bibliography{Primordial_Modes}






\end{document}